\documentclass[
12pt,%
a4paper,%
notitlepage,%
oneside,%
onecolumn,%
nobibnotes,%
nofootinbib,%
noshowpacs]{article}
\begin{document}
\title{Geoneutrino and Hydridic Earth model. Version 2}
\author{L.Bezrukov$^a$}
\maketitle
$^a$ Institute for Nuclear Research of Russian academy of sciences, Moscow

\begin{abstract}
$U, Th$ and $K^{40}$ abundances in the Earth were calculated 
in the frame of Hydridic Earth model. Terrestrial heat producton from $U, Th$ and $K^{40}$ decays was calculated also. We must admit the existance of Earth expansion process to understand the obtained large value of terrestrial heat producton. The geoneutrino detector with volume more than 5 kT (LENA type) must be constructed to definitely separate between Bulk Silicat Earth model  and Hydridic Earth model.  In second version of the article we assume that $K^{40}$ concentration distributes in the Earth uniformly.
\end{abstract} 
\section{Introduction.}
\hspace{0.5cm}Geoneutrino is antineutrino emitted in a decay chain of $U,Th$ and $K^{40}$ located in the Earth's interior. The experimental information on the geoneutrino fluxes and energy distribution can help to build a realistic Earth model. The first direct measurement of geoneutrino flux was made by the Borexino collaboration \cite{Borex2013} and the KamLand collaboration \cite{KamLand2013}.  Number of events in these detectors depends on the uranium mass in the Earth $m(U)$, the thorium mass in the Earth $m(Th)$ and on their distribution in the Earth.  The Bulk Silicat Earth (BSE) model \cite{Mantovani2003} gives  $m_{BSE}(U) = 0,81 \cdot 10^{17}kg, m_{BSE}(Th) = 3,16 \cdot 10^{17}kg, m_{BSE}(K) = 0.49 \cdot 10^{21}kg $. This amount distributes only in Crust and Upper Mantel in the frame of BSE model.

There is the alternative Earth model \cite{Larin}, \cite{Larin2012} named Hydridic Earth model which predicts the primordial chemical elements composition of the Earth. 

Vladimir Larin \cite{Larin} used the idea that the separation of the chemical elements in the solar system (chemical differentiation) was originated from the magnetic
field of the Protosun. He found a correlation between the ratio of the Earth crust chemical element abundances to Sun chemical
element abundances and the first ionization potential of these elements.
The observed correlation is theoretically \cite{Larin2012} interpreted as a Boltzmann distribution. The numerical model was succesfully tested for the observed solar normalized chemical compositions of the Earth, Mars and chondrites. 

The 18.3\% of the Earth primordial mass is predicted to be Hydrogen  \cite{Larin2012}. 
The inner Earth would have been and still
could be hydrogen rich. The most part of primordial hydrogen have
escaped to atmosphere and space through the degassing of the mantle. Model suggests that large amounts of hydrogen are still 
located in the core.

\vspace{0.6cm}
\begin{tabular}{lll}
\multicolumn{3}{c}{\bf Geochemical model of modern Hydridic Earth}\\
Geosphere & Depth range, km & Composition \\
\bf Lithosphere & 0 - 150 & $CaO, MgO, Al_{2}O_{3}, SiO_{2}, Na_{2}O,$\\
&&$FeO, H_{2}O ...$ \\
\bf Astenosphere & 150 & Thin layer of Metalsphere\\
&&with high hydrogen concentration\\
\bf Metalsphere & 150 - 2900 &  $Mg_{2}Si : Si : FeSi =6 : 3 : 1$\\

\bf External core & 2900 - 5000 & $MgH_{0.1}, SiH_{0.1}, FeH_{0.1}$ \\
& & $+ MgH_{n}, SiH_{n}, FeH_{n}$  (n = 10) \\

\bf Inernal core & 5000 - 6371 & $MgH_{n}, SiH_{n}, FeH_{n}$  (n = 10) \\
\end{tabular}

\section{Uranium abundance.}

\hspace{0.5cm}The law of magnetic chemical differentiation of planets proposed by \cite{Larin2012} gives the following expression for abundance of chemical element with atomic number M in the Earth:

\begin{equation}\label{eq:1}
\frac{(\frac{X_{M}}{X_{Si}})_{Earth}}{(\frac{X_M}{X_{Si}})_{Sun}}
= 12108 \cdot e^{-1.1537 \cdot E_{IP}(M)}, 
\end{equation}
where $X_{M}$ is the mass fracton in the planet mass of the chemical element with atomic number M,
$E_{IP}(M)$ is the ionization potential of the chemical element with atomic number M in eV. The $E_{IP}(U) = 6.2 eV$ for uranium and from (\ref{eq:1}) we have:

\begin{equation}\label{eq:2}
(\frac{X_{U}}{X_{Si}})_{Earth} = (\frac{X_U}{X_{Si}})_{Sun} \cdot 9.475.
\end{equation}

We shall use this expression to calculate U abundance $X_U$ in the Earth. For that we need to know $(\frac{X_U}{X_{Si}})_{Sun}$ for the Sun. There are no experimental data \cite{Asplund} for the Sun but for Asteroid Belt (AB) such data exist \cite{Asplund}. 

The expression for abundance of chemical element M in the AB according to \cite{Larin2012} has the following form: 

\begin{equation}\label{eq:3}
\frac{(\frac{X_{M}}{X_{Si}})_{AB}}{(\frac{X_M}{X_{Si}})_{Sun}}
= 2.8425 \cdot e^{- 0.1282 \cdot E_{IP}(M)}, 
\end{equation}

From this formula we can calculate:

\begin{equation}\label{eq:4}
(\frac{X_{U}}{X_{Si}})_{Sun} = (\frac{X_U}{X_{Si}})_{AB} \cdot \frac{1}{1.284}.
\end{equation}

 To calculate  $(\frac{X_U}{X_{Si}})_{AB}$ we shall take the experimental values of $\lg \epsilon_U - \lg \epsilon_{Si}$ from \cite{Asplund} where $\epsilon_U$ is the number of $U$ atoms containing in chondrits in unit of volume:

\begin{equation}\label{eq:5}
(\frac{X_U}{X_{Si}})_{AB}
= \frac{238}{28} \cdot 10^{(\lg \epsilon_U - \lg \epsilon_{Si})} = 7.93 \cdot 10^{- 8}.
\end{equation}

Substituting the result from (\ref{eq:5}) in (\ref{eq:4}) we obtain
 $(\frac{X_{U}}{X_{Si}})_{Sun} = 6.18 \cdot 10^{- 8}$. 
Substituting this value in (\ref{eq:2}) we obtain

\begin{equation}\label{eq:6}
(\frac{X_U}{X_{Si}})_{Earth} = 5.85 \cdot 10^{- 7}.
\end{equation}

  The mass fracton in the Earth mass of silicon from \cite{Larin2012} is $X_{Si} = 9,028 \cdot 10^{- 2}$.  We can calculate from (\ref{eq:6}) the uranium mass in the present day Earth knowing the Earth mass  $m_{Earth} = 5.97 \cdot 10^{24}kg$ :

\begin{equation}\label{eq:7}
m(U) = X_{Si} \cdot m_{Earth} \cdot 5.85 \cdot 10^{- 7} = 3.15 \cdot 10^{17} kg.
\end{equation}

 Compare this value with  $m_{BSE}(U) = 0,81 \cdot 10^{17}kg$.

We used the present day experimental data for the Sun and AB, so there is no need to take into account the $U$ mass loss due to $U$ decay.

\section{Thorium abundance.}

\hspace{0.5cm}Here we use the same logic as in the previous section. The $E_{IP}(Th) = 6.95 eV$ for thorium  and from (\ref{eq:1}) we have:

\begin{equation}\label{eq:8}
(\frac{X_{Th}}{X_{Si}})_{Earth} = (\frac{X_{Th}}{X_{Si}})_{Sun} \cdot 3.988.
\end{equation}

From (\ref{eq:3}) we have:

\begin{equation}\label{eq:9}
(\frac{X_{Th}}{X_{Si}})_{Sun} = (\frac{X_{Th}}{X_{Si}})_{AB} \cdot \frac{1}{1.165}.
\end{equation}

From (\ref{eq:5}) by using experimental data for AB from \cite{Asplund} we  have:

\begin{equation}\label{eq:10}
(\frac{X_{Th}}{X_{Si}})_{AB}
= \frac{232}{28} \cdot 10^{(\lg \epsilon_{Th} - \lg \epsilon_{Si})} = 2.94 \cdot 10^{- 7},
\end{equation}

From  (\ref{eq:9}, \ref{eq:8}) we have:

\begin{equation}\label{eq:11}
(\frac{X_{Th}}{X_{Si}})_{Sun} = 2.52 \cdot 10^{- 7},
\end{equation}

\begin{equation}\label{eq:12}
(\frac{X_{Th}}{X_{Si}})_{Earth} = 1.0 \cdot 10^{- 6}.
\end{equation}

So we can calculate the present day thorium mass in the Earth:

\begin{equation}\label{eq:13}
m(Th) = X_{Si} \cdot m_{Earth} \cdot 1.0 \cdot 10^{- 6} = 5.42 \cdot 10^{17} kg.
\end{equation}

 Compare this value with $m_{BSE}(Th) = 3,16 \cdot 10^{17}kg$.

\section{Potassium-40 abundance.}

\hspace{0.5cm} The logic for potassium is simple because $X_K = 3.76 \cdot 10^{- 2}$ for the Earth is known from \cite{Larin2012} and fraction $\frac{X_{K40}}{X_{K}} = 1.17 \cdot 10^{- 4}$ for present day Earth is known also.

So, we have for potassium-40 mass in the Earth:
\begin{equation}\label{eq:14}
m(K^{40}) = X_{K} \cdot m_{Earth} \cdot 1.17 \cdot 10^{- 4} = 2.63 \cdot 10^{19} kg.
\end{equation}
 
 Compare this value with

 $m_{BSE}(K^{40}) = m_{BSE}(K) \cdot 1.17 \cdot 10^{- 4} = 5.73 \cdot 10^{16}kg$.

Hydridic Earth model predicts the huge amount of $K^{40}$ in the Earth.

\section{$Th/U$ mass ratio.}

\hspace{0.5cm} $Th/U$ mass ratio is defined as $R = \frac{m_{Th}}{m_{U}}$. 

From (\ref{eq:7}) and (\ref{eq:13}) we have the Hydridic Earth model prediction for the Earth:
\begin{equation}\label{eq:15}
R_{Earth} = (\frac{m_{Th}}{m_{U}})_{Earth} = 1.72.
\end{equation}

From (\ref{eq:9}) and (\ref{eq:4}) we have the Hydridic Earth model prediction for the Sun:
\begin{equation}\label{eq:16}
R_{Sun} = (\frac{m_{Th}}{m_{U}})_{Sun} = 4.09.
\end{equation}

From (\ref{eq:5}) and (\ref{eq:10}) we have the value measured for the Asteroid Belt:
\begin{equation}\label{eq:17}
R_{AB} = (\frac{m_{Th}}{m_{U}})_{AB} = 3.71.
\end{equation}

The fit to Borexino geoneutrino data \cite{Borex2013} was made with $m(U)$ and $m(Th)$ as a free parameters. The following experimental values of signal were obtained:

 $S_{Th} = (10.6 \pm 12.7)$TNU,

 $S_{U} = (26.5 \pm 19.5)$TNU. 

Let us calculate the ratio of $S_{Th}$ to $S_{U}$ using these experimental values:
\begin{equation}\label{eq:18}
(\frac{S_{Th}}{S_U})_{exp} = 0.4 \pm 0.78. 
\end{equation}

This ratio was theoretically calculated for $R_{AB} = 3.9$ in \cite{Borex2013}:

\begin{equation}\label{eq:19}
(\frac{S_{Th}}{S_U})_{BSE} = 0.25.
\end{equation}

So we can calculate this ratio for our value $R_{Earth} = 1.72$ tacking into account that  this ratio is proportional to $R_{Earth}$:

\begin{equation}\label{eq:20}
(\frac{S_{Th}}{S_U})_{HydridicEarth} = 0.25 \cdot \frac{1.72}{3.9} = 0.11.
\end{equation}

The both values (\ref{eq:19}) and (\ref{eq:20}) are consistent with (\ref{eq:18}). So, the prediction of the Hydridic Earth model (\ref{eq:15}) does not contradict to Borexino experimental data.
It is reasonable to analyse the Borexino data assuming a fixed mass $Th/U$ ratio of 1.72.

The accurate measurement of $Th/U$ ratio could permit to choose between BSE model and Hydridic Earth model. 
The low background geoneutrino detector with volume more than 5 kT (LENA type) must be constructed for that purpose \cite{Sinev2010},\cite{Barabanov2009} in the place far from nuclear reactors.

\section{Terrestrial heat production.}

\hspace{0.5cm}In the Hydridic Earth model the total masses of uranium, thorium and potassium-40 are (\ref{eq:7}, \ref{eq:13}, \ref{eq:14})
 $m(U) = 3.15 \cdot 10^{17} kg$,
$m(Th) = 5.42 \cdot 10^{17} kg$, 
$m(K^{40}) = 2.63 \cdot 10^{19} kg$.

The equation relating masses and heat production is
\begin{equation}\label{eq:21}
H = m \cdot \frac{N_{Avog}}{A} \cdot \frac{E_{release}}{\tau} \cdot \alpha, 
\end{equation}
	where $N_{Avog}$ - Avogadro number, $A$ - atomic number, $E_{release}$ - energy release in decay chain in MeV, $\tau = t_{1/2} / ln2$ - mean lifetime of isotope, $\alpha$ - the conversion factor $1 MeV = 1.602 \cdot 10^{-13} J$
\begin{equation}\label{eq:22}
H(U) = 3.15 \cdot 10^{20} g \frac{6 \cdot 10^{23}}{238} g^{-1} \cdot \frac{47.6 MeV}{6.45 \cdot 10^9 \cdot 3.15 \cdot 10^7 s} \cdot 1.602 \cdot 10^{-13} J = 29.8 TW. 
\end{equation}
\begin{equation}\label{eq:23}
H(Th) = 5.42 \cdot 10^{20} g \frac{6 \cdot 10^{23}}{232} g^{-1} \cdot \frac{41.5 MeV}{2.03 \cdot 10^{10} \cdot 3.15 \cdot 10^7 s} \cdot 1.602 \cdot 10^{-13} J = 14.6TW. 
\end{equation}
\begin{equation}\label{eq:24}
H(K40) = 2.63\cdot 10^{22} g \frac{6 \cdot 10^{23}}{40} g^{-1} \cdot \frac{0.6 MeV}{1.8 \cdot 10^9 \cdot 3.15 \cdot 10^7 s} \cdot 1.602 \cdot 10^{-13} J = 669 TW. 
\end{equation}
\begin{equation}\label{eq:25}
H = H(U) + H(Th) + H(K40) = 713 TW. 
\end{equation}

The see that the active geo-reactor in the Earth core is not necessary to explain the observed power flux going up $H_{observed} = 44 TW$ \cite{Pollack1993} on the Earth surface. On the contrary the obtained heat production is too high. This heat is produced mostly by $K^{40}$ decay.

\section{$U$ and $Th$  distribution in the Earth.}

\hspace{0.5cm}The experimental signal of the geoneutrino detectors depends not only on $m(U)$ and $m(Th)$ but also on the distribution of $U$ and $Th$ in the Earth's  interior. 

The calculations show that decays in the crust matter surrounding of the detector give the main contribution to the experimental signal \cite{Mantovani2003}. 
The experimental data from Borexino consistent with BSE model with $m_{BSE}(U) = 0,81 \cdot 10^{17}kg$ distributed in the crust and upper mantel of the Earth. 
Taking into account these facts and the prediction of the Hydridic Earth model  (\ref{eq:7}, \ref{eq:20}) we can obtain that values

\begin{equation}\label{eq:26}
 m_{Lithosphere}(U) = 1 \cdot 10^{17} kg, \hspace{1cm}
m_{core}(U) = 2.15 \cdot 10^{17} kg 
\end{equation}
are not contradict to Borexino results.

Using  (\ref{eq:15}) and  (\ref{eq:13}) we can obtain for $Th$ distribution: 
\begin{equation}\label{eq:27}
 m_{Lithosphere}(Th) = 1.7 \cdot 10^{17} kg, \hspace{1cm}
m_{core}(Th) = 3.7 \cdot 10^{17} kg.
\end{equation}
The Hydridic Earth model predicts the existence of
the Earth hydrogen degassation process. Hydrogen appears on the surface of the Earth core and goes up to the cosmic space through the long chain of processes. During this way the hydrogen flow purifies the mantel volume up to lithosphere. So $U$ and $Th$ must locate in the present-day Earth mostly in the core and the lithosphere.

The Hydridic Earth model predicts also
the expansion of the Earth. The model predicts the increasing of Earth radius in 1.71 time during Earth life \cite{Larin} from the moment of origing to present days. In the frame of this picture the primordial Earth radius was $R_{prim} = 6371/1.71 km = 3726 km$.
The Internal core of present-day Earth still could be primordial and could have the primodial $U$ and $Th$  densities. The Internal core of present-day Earth density predicted by Hydridic model is $\rho_{In} = 25 g \cdot cm^3$.  The External core of present-day Earth is ranged from the depth $R_{Ex} = 2900 km$ to $R_{ In} = 5000 km$ with density  $\rho_{Ex} = 12 g \cdot cm^3$. So, the existence in the present day Earth core of the large amount of $U$ and $Th$  is natural in the frame of the  Hydridic Earth model.

\section{ Potassium distribution in the Earth and terrestrial heat flux.}

\hspace{0.5cm}Unfortunately the Borexino and Kamland detectors can not measure the antineutrino from $K^{40}$ decay and we can not say here something definitly about $K$ distribution in the Earth from the data of these experiments. 
 Let assume  that $K$ concentration is distributed uniformly in the Earth and equal $\rho_K = 3.76 \cdot 10^{- 2}g/g$. 

So, the Hydridic Earth model predicts the following heat producton in the Earth lithosphere and metalsphere:

\begin{equation}\label{eq:28}
H_{Lithosphere} = (H_{U} + H_{Th})/3 = 15 TW.
\end{equation} 
\begin{equation}\label{eq:29}
H_{metalsphere+core}  = 698 TW.
\end{equation} 

These figures (\ref{eq:28}, \ref{eq:29}) show that the heat must flow from metalsphere to the space mostly through the oceans where the crust is more thin than continental one.  The existing experimental data support this statement. 

The second important idea introduced in \cite{Larin} is that the intrinsic Earth heat can be absorbed during Earth expansion period by the hydrid decomposition. This idea has the interesting consequence: the Earth heat flux measured near Earth surface can be not stable. Before the period of the Earth expansion the flux is increased  and after the finishing of the Earth expansion perod the flux is decreased.

Here we refer to Argo project \cite{Argo} results.  Argo consists of approximately 8200 small (20-30 kg) drifting robotic probes deployed worldwide. The Argo measured the ocean temperature at the depths up to 2000 m during 2005-2010 years. This period was the solar minimum but Argo recorded the increasing of the ocean waters temperatute or the oceanic heat content (Oceanic heat content is the heat stored in the ocean).  Taking into account this Argo result the work \cite{Hansen} shows that the Earth energy imbalance  is $0.58 \pm 0.15 W m^{-2}$ during the 6-yr period 2005-2010. Let multiply this value of energy imbalance on the value of Earth surface:
\begin{equation}\label{eq:30}
0,58 W m^{-2}  \cdot  5,1 \cdot 10^{14} m^2 = 300 \pm 76 TW
\end{equation}
and compare with (\ref{eq:29}). So, the obtained value of the heat production in the metalsphere and the core is enough to explain the value of  global oceanic heat content measured by Argo project. The increasing of the radiogenic intrinsic Earth heat flux can be considered as a source of increasing of global oceanic heat content. 

But this idea has the difficuties. Firstly the other ideas can be proposed to explain the increasing of oceanic heat content. Secondly the obtained here value of heat production (\ref{eq:25}) contradicts to the value measured for the Earth \cite{Pollack1993} and equal to 44TW. 
	
How we can investigate this problem experimentally? We propose that  the precision measurements of heat flux in the crust during long period  are highly important to investigate the flux stability. 

 \section{ $K^{40}$ antineutrino flux.}

\hspace{0.5cm}Let estimate the $K^{40}$ antineutrino flux $F_{\tilde{\nu}}(K^{40})$ on the Earth surface. The energy spectrum of $K^{40}$ antineutrino has $E_{\tilde{\nu}}^{max} = 1,31 MeV$. Suppose for simplicity that all $K^{40}$ mass is located in the Earth center.
\begin{equation}\label{eq:31}
F_{\tilde{\nu}}(K^{40}) > \frac{m(K^{40}) \cdot N_{Avog} \cdot \xi}{A \cdot \tau(K^{40})   \cdot 4 \pi \cdot R_{Earth}^2} = 7.8 \cdot 10^8 cm^{-2} s^{-1}, 
\end{equation}
where $\xi = 0 .45 $ is the factor taking into account the oscillation of antineutrino and the fact that only part of $K^{40}$ decays emits antineutrino.

Compare this value with the fluxes of $Be^7$ solar neutrino with energy $E_{\nu} = 0,86 MeV$ and $pep$ solar neutrino with energy $E_{\nu} = 1.44 MeV$ on the Earth predicted by Standart Solar Model and neutrino oscillations:

$ F_{\nu}(Be^7) = 4.8 \cdot 10^9 cm^{-2} s^{-2}$,

$ F_{\nu}(pep) = 1.42 \cdot 10^8 cm^{-2} s^{-2}$.

The scintillator detector can record $\nu$ and $\tilde{\nu}$ at these energies via reactions:
\begin{equation}\label{eq:32}
\tilde{\nu} + e = \tilde{\nu}' + e'
\end{equation}
\begin{equation}\label{eq:33}
\nu + e = \nu' + e'
\end{equation}

To compare the number of electron events induced by reaction(\ref{eq:32}) and  (\ref{eq:33}) we must take into account that energy distribution of recoil electrons in reaction  (\ref{eq:32}) is different from reaction  (\ref{eq:33}). The $\tilde{\nu}$ contribute mostly to event with low electron energy compareing $\nu$ contribution. The accurate calculation of the $K^{40}$ induced events in Borexino \cite{Borex2007}, \cite{Borex2011} and Kamland backgrounds is needed.

\section{Conclusion.}

\hspace{0.5cm}1. On the base of Hydridic Earth model we calculated $U, Th$ and $K^{40}$ abundances in the Earth:
(\ref{eq:7}) $m(U) = 3.15 \cdot 10^{17} kg$,  (\ref{eq:13})  $m(Th) = 5.42 \cdot 10^{17} kg$  (\ref{eq:14})  $m(K^{40}) = 2.63 \cdot 10^{19} kg$.

2. The obtained $Th/U$ mass ratio for the Earth  $(\frac{m_{Th}}{m_{U}})_{Earth} = 1.72$ is different from chondritic $Th/U$ mass ratio of 3.9 usually used. This predicted by Hydridic Earth model value does not contradict to Borexino geoneutrino data. The accurate measurement of this ratio could permit to choose between BSE model and Hydridic Earth model. The geoneutrino detector with volume more than 5 kT (LENA type) must be constructed for that purpose \cite{Sinev2010},\cite{Barabanov2009}.  

3. In the frame of Hydridic Earth model the part of $U$ and $Th$  mass are contained in the Earth core. We estimated this part by using Borexino data.

4. We calculated the heat production in the Earth induced by $U, Th$ and $K^{40}$ decays. Total heat production is $H = H(U) + H(Th) + H(K^{40}) = 713 TW$.  The $K^{40}$ decays produce the most part of this heat. We showed that in the frame of the Hydridic Earth model the intrinsic Earth heat must flow  mostly through the oceans where the crust is more thin.  The obtained value of heat production in the metalsphere and the core is enough to explain the value of  global oceanic heat content measured by Argo project. The increasing of the radiogenic intrinsic Earth heat flux can be considered as a source of increasing of global oceanic heat content.

5. In the frame of Hydridic Earth model the Earth heat flux measured near Earth surface can be not stable. Before the period of the Earth expansion the flux is increased  and after the finishing of the Earth expansion perod the flux is decreased.

The precision measurements of the heat flux in the crust during long period are highly important  to investigate the flux stability.

6. We estimated the $K^{40}$ antineutrino flux on the Earth surface. This flux is compearable with $Be^7$ and $pep$ solar neutrino fluxes on the Earth surface. So the accurate calculation of the $K^{40}$  induced events in Borexino \cite{Borex2007}, \cite{Borex2011} and Kamland backgrounds is needed.

\section{Acknowledgements.}

\hspace{0.5cm}I am grateful to Vladimir Larin, Bayarto Lubsandorzhiev, Stefan Schoenert and Valery Sinev for useful discussions.
This work was supported by grant of Russian Foundation of Basic Research No 13-02-92440.


\end{document}